\documentclass[reprint,amsmath,amssymb,aps,prl]{revtex4-2}
\usepackage{graphicx}  
\usepackage{bm}        
\usepackage{amssymb}   
\usepackage{dcolumn}
\usepackage{bm}
\usepackage{mathtools}
\usepackage{dsfont}

\newcommand{\uxi}{u(x,\xi)}
\newcommand{\uxp}{u(x,\xi')}
\newcommand{\st}{\mu_t}
\newcommand{\sa}{\mu_a}
\newcommand{\sct}{\frac{\mu_s}{2}}
\newcommand{\uvi}{u(v,\xi,t)}
\newcommand{\uvp}{u(v,\xi',t)}

\begin{document}
\widetext

\title{Fine Boundary--Layer structure in Linear Transport Theory}
\author{E.~L.~Gaggioli}
\email{egaggioli@iafe.uba.ar}
\affiliation{Instituto de Astronom\'{\i}a y F\'{\i}sica del Espacio
(IAFE, CONICET--UBA), Casilla de Correo 67 - Suc. 28 (C1428ZAA), Buenos Aires, Argentina.}
\author{D.~M.~Mitnik}
\affiliation{Instituto de Astronom\'{\i}a y F\'{\i}sica del Espacio
(IAFE, CONICET--UBA), Casilla de Correo 67 - Suc. 28 (C1428ZAA), Buenos Aires, Argentina.}
\author{O.~P.~Bruno}
\affiliation{Computing and Mathematical Sciences, Caltech, Pasadena, CA 91125, USA.}

\begin{abstract}
  This contribution identifies and characterizes a previously
  unrecognized boundary--layer structure that occurs in the context of
  linear transport theory, with an impact on the fields of Radiative
  Transfer and Neutron Transport. The existence of this boundary layer
  structure, which governs the interactions between different
  materials, or between a material and vacuum, plays a critical role
  in a correct description of transport phenomena. Additionally, the
  boundary--layer phenomenology explains and helps bypass
  computational difficulties reported in the literature over the last
  several decades.
\end{abstract}

\maketitle

The linear transport equation provides a general model for neutral
particle transport, including \textit{e.g.}~Neutron Transport and Radiative
Transfer of photons, which impacts upon a wide range of disciplines in
science and technology
\cite{Chandrasekhar1960,Thynell1998,Mishchenko1999,Vassiliev2010,Qin2015}. In
this context, this paper identifies and characterizes a significant
boundary--layer structure which, albeit present in \textit{e.g.}~eigenfunction 
systems for one dimensional
problems~\cite[Ch. 4]{Case1967} (as they must be in view of the
theoretical discussion presented in this paper), has not previously
been correctly accounted for or even fully recognized, namely, a structure of sharp
solution transitions with unbounded gradients arbitrarily close to
interface boundaries.
\begin{figure}[h!]
\centering
  \includegraphics[width=0.85\linewidth]{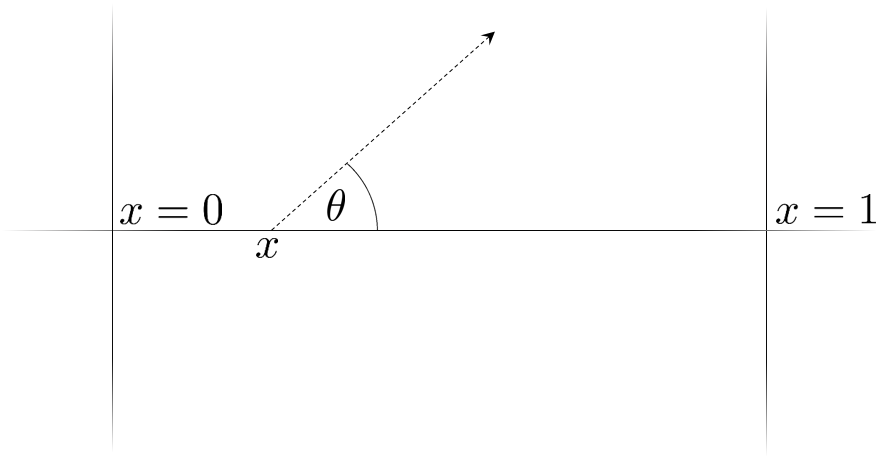}
  \caption{One--dimensional ``slab'' geometry: $\xi = \cos(\theta)$.}
 \label{fig:parallelgeom}
\end{figure}
The equation for time--independent neutral
particle transport in a one--dimensional plane--parallel geometry
(Fig.~\ref{fig:parallelgeom}), with isotropic scattering and Fresnel
boundary conditions, is given by
\begin{equation}
\begin{split}
  &\xi \frac{\partial}{\partial x}\uxi+\mu_t(x)\uxi= \\
  &\qquad \;\;\;\;\; \frac{\mu_s(x)}{2} \int_{-1}^{1}\uxp d\xi' +q(x,\xi),\\
  &u(x=0,\xi)=\mathcal{R}(\xi)u(x=0,\xi_R) \; \forall \xi>0,\\
  &u(x=1,\xi)=\mathcal{R}(\xi)u(x=1,\xi_R) \; \forall \xi<0.
\end{split}
\label{eq:transp1d}
\end{equation}
Here $\xi=\cos(\theta)$, and, letting $\mu_s(x)$ and $\mu_a(x)$ denote
the macroscopic scattering and absorption coefficients, respectively,
the total transport coefficient is $\st(x)=\sa(x)+\mu_s(x)$. The
integral term accounts for the angular redistribution of
particles due to scattering,  $q(x,\xi)$ is an external source, and 
the Fresnel boundary conditions model the reflection of
particles at the boundary interface due to refractive-index
differences---wherein $\xi_R=-\xi$ equals the reflected direction cosine
and $\mathcal{R}(\xi)$ the corresponding Fresnel reflection
coefficient~\cite{Born1999}.

Noting that the coefficient $\xi$ of the highest order derivative
in~\eqref{eq:transp1d} tends to zero as $\theta \to \pi/2$, the
existence of sharp spatial gradients in the solution $u(x,\xi)$ for
such angles and for $x$ near the interfaces $x=0$ and $x=1$ may be
expected~\cite[Ch. 9]{Bender1999}, leading to  unbounded spatial gradients
at points $(x,\xi)$ near $(0,0)$ and $(1,0)$. Such
``boundary--layer'' structures, which are caused by the existence of a
spatial boundary condition in conjunction with a vanishingly small
coefficient for the highest order differential operator, only take
place for the incoming directions $\xi>0$ (resp. $\xi<0$) for points
close to $x=0$ (resp. $x=1$)---since it is for such directions that
boundary conditions are imposed in Eq.~\eqref{eq:transp1d}.

Following~\cite{Bender1999}, in order to characterize the boundary
layer structure around \textit{e.g.} $x=0$, an {\em inner solution}
of the form $U(X,\xi) = u(\xi X,\xi)$ is employed; the boundary layer around $x=1$
can be treated analogously. The lowest--order asymptotics $U_0(X,\xi)$
of $U(X,\xi)$ as $\xi\to 0^+$ satisfies the {\em constant coefficient}
equation
\begin{equation}
\begin{split}
\frac{\partial U_0(X,\xi)}{\partial X}& + \mu_t(0) U_0(X,\xi)=\\
&\qquad \frac{\mu_s(0)}{2} \int_{-1}^{1} U_0\left(\frac{\xi X}{\xi'},\xi'\right) d\xi' +q(0,\xi),\\
\end{split}
\label{eq:transp1d2}
\end{equation}
and boundary conditions induced by~\eqref{eq:transp1d}.
Eq.~\eqref{eq:transp1d2} tells us that the derivative
$\frac{\partial U_0(X,\xi)}{\partial X}$ remains bounded as
$\xi\to 0^+$, and, therefore, that the function $U_0(X,\xi)$
characterizes the boundary layer structure in the solution $u(x,\xi)$
via the relation
\begin{equation}
  u(x,\xi)\sim u_0(x,\xi) = U_0(x/\xi,\xi) \; \text{as}  \;  (x,\xi)\to (0^+,0^+).
\end{equation}
The argument can be extended to two and three--dimensional problems,
and to include time--dependence and curved boundaries. In such cases a
boundary layer occurs with large solution derivatives near the
boundary in directions parallel to the domain interface.

Utilizing an integrating factor and letting
\begin{equation*}
I(x,\xi) = \int_0^{x} e^{\frac{ \mu_t(0) y}{\xi}}\left[\frac{\mu_s(0)}{2} \int_{-1}^1u_0(y,\xi')d\xi'+q(0,\xi)\right]dy,
\end{equation*}
the boundary layer solution 
\begin{equation}
 u_0(x,\xi) = \frac{e^{-\mu_t(0)x/\xi}}{\xi} \Big[\xi u(0,\xi)+I(x,\xi)\Big] ,
 \label{eq:intf}
\end{equation}
is obtained, which explicitly exhibits the
exponential boundary--layer character of the solution. 

The boundary layer structure can be visualized by considering
the exact solution of Eq.~\eqref{eq:transp1d} that results for
the scattering--free case ($\mu_s(x)=0$) with constant--coefficients. Then 
\begin{equation}
\begin{split}
u(x,\xi)=\begin{cases}
\displaystyle \frac{q}{\mu_a}\left[1\;-\;\;\frac{\eta(\xi)}{ e^{\mu_a x / \xi}}\;\;\;\right]&\forall\xi>0,\\[8pt]
\displaystyle  \frac{q}{\mu_a}\left[1- \frac{\eta(\xi)}{e^{\mu_a (x-1) / \xi}} \right]&\forall\xi<0,
\end{cases}
\end{split}
\label{eq:transpnscatt}
\end{equation}
results, where
\begin{equation*}
\;\;\eta(\xi)=\frac{\mathcal{R}(|\xi|)-1}{\mathcal{R}(|\xi|)e^{-\mu_a/|\xi|}-1}.\qquad\qquad\qquad
\end{equation*}
For simplicity only vacuum boundary conditions ($\mathcal{R}(\xi)=0$) are considered in what follows. The general case can be treated analogously.
\begin{figure}[h!]
\centering
  \includegraphics[width=0.8\linewidth]{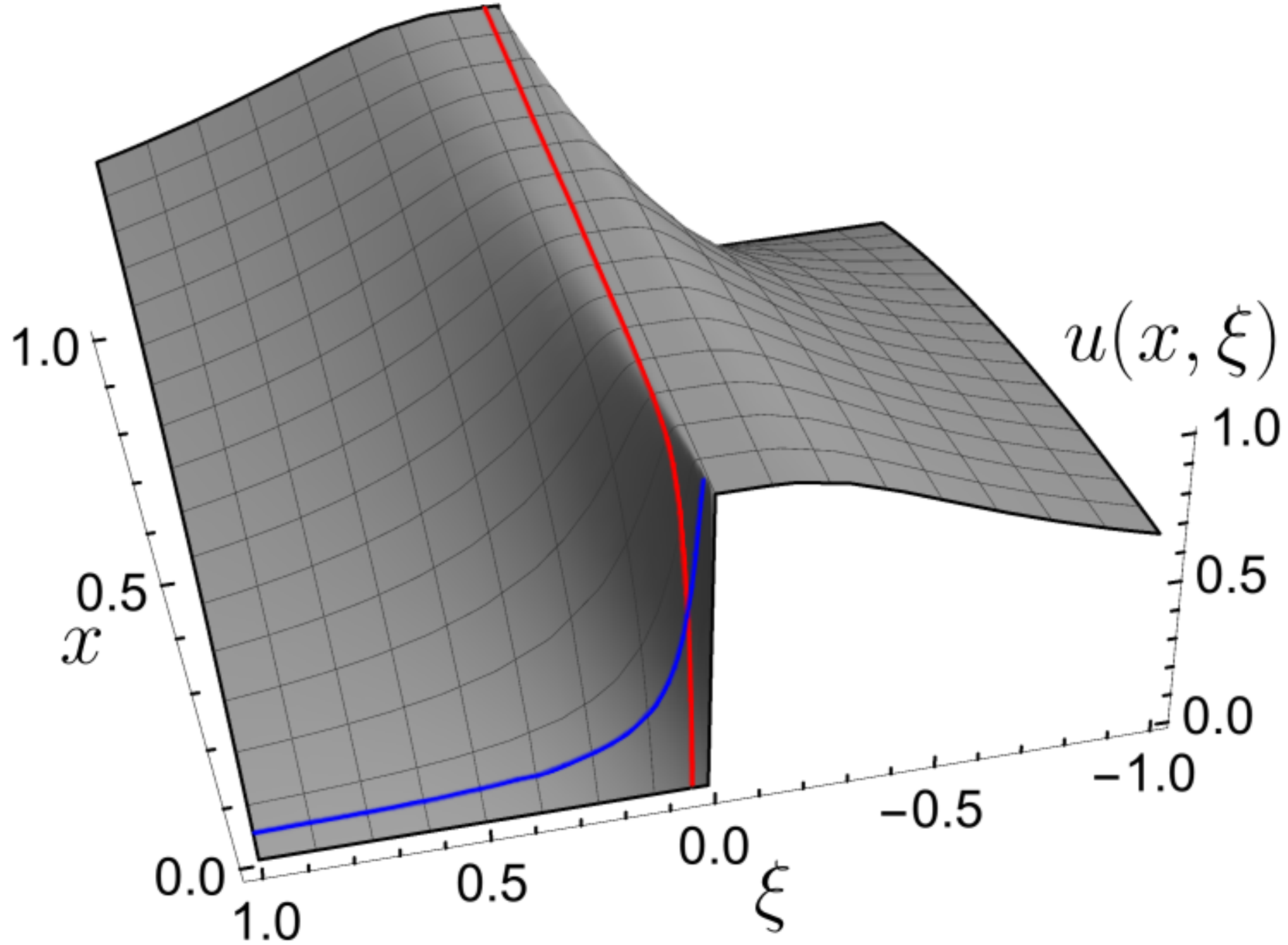}
  \caption{Solution $\uxi$ of Eq.~\eqref{eq:transp1d} given by Eq.~\eqref{eq:transpnscatt} with 
  $\mathcal{R}(\xi)=0$. The boundary layer in the $x$ and $\xi$ 
  variables is clearly visible in the blue and red superimposed curves.}
 \label{fig:ansol}
\end{figure}
Fig.~\ref{fig:ansol} clearly displays the boundary layer structure 
as $(x,\xi) \to (0^+,0^+)$. 

The foremost two numerical
methods used in the area, namely, the Spherical Harmonics Method and
the Discrete Ordinates Method, see \textit{e.g.}~\cite[Ch. 8]{Case1967}
and~\cite[Ch. 3, 4]{Lewis1984}, fail to
properly resolve such boundary layer structures, 
as shown by recent attempts \cite{Rocheleau2020,Wang2019,Harel2020,Anli2006,Chai1993}. 

In view of the expression~\cite[p. 77]{Trefethen2008} for the error
term in Gauss--Legendre integration, the $\ell$--point quadrature
error decreases like $32V/15\pi j (2\ell+1-j)^j$ provided
the $j \leq 2\ell$ derivative of the integrated function is bounded by
$V$. Introducing the change of variables $\xi'=r^p$ in the integral
in~\eqref{eq:transp1d}
we thus seek a bound $V$ on the $j$--th derivative of the resulting
integrand. Using an integrated version of~\eqref{eq:transp1d}, similar
to~\eqref{eq:intf}, combining two exponential terms and using
the fact that for each non-negative integer $k$ the integral $\int_0^\infty t^k e^{-t} dt$ is finite, we find that
\begin{equation*}
\begin{split}
\left | \frac{\partial^j }{\partial r^j}\left[ u(x,r^p) r^{p-1}\right]\right|\leq W r^{p-j-1}\; \text{as}  \; (x,r) \to (0^+,0^+),
\end{split}
\label{eq:boundedder}
\end{equation*}
for some constant $W$; setting $V = W r^{p-j-1}$ yields the desired
bound, which, importantly, {\em is uniform for all relevant
  values of $x$ and $r$} (as long as $p \ge j+1$).

In the case $\mathcal{R}(\xi)=0$ considered presently, splitting the
integral on the right hand side of Eq.~\eqref{eq:transp1d} at the
boundary-layer point $\xi=0$ yields
\begin{equation}
\int_{0}^{1} u(x,\xi) d\xi \sim \sum_{i=1}^{M/2} w_i u(x,\xi_i),
\label{eq:MartensenKussmaul}
\end{equation}
where, letting $r_i$ and $w_i^{GL}$ denote the Gauss--Legendre
quadrature abscissas and weights in the interval $[0,1]$, we have set
$\xi_i=r_i^p$ and $w_i= p \times r_i^{p-1} \times w_i^{GL}/2$. The
abscissas and weights corresponding to the negative values of $\xi$
are obtained by symmetry.
\begin{figure}[h!]
\centering
  \includegraphics[width=0.85\linewidth]{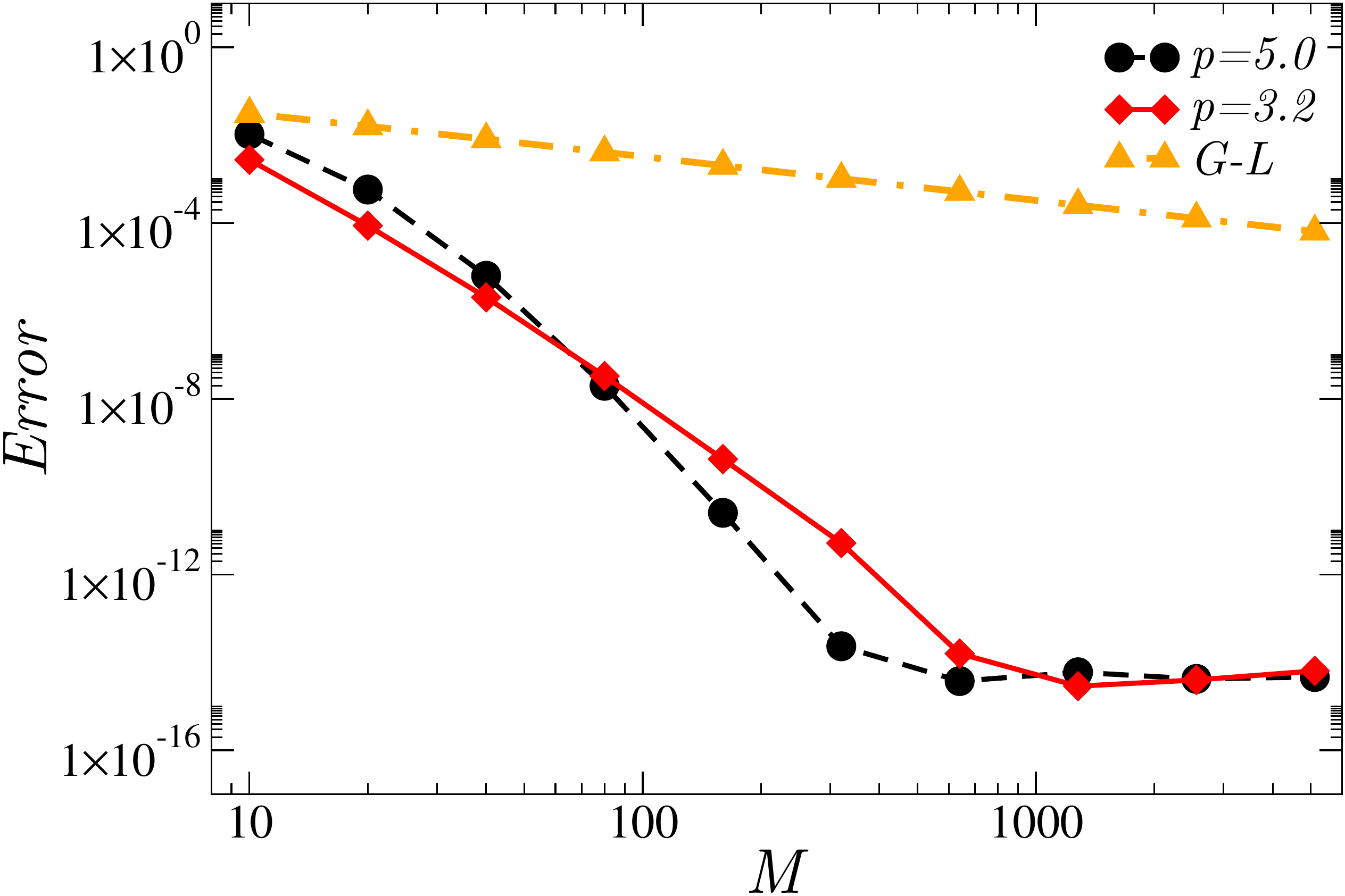}
  \caption{Error of numerical against analytic integral,
    $E(M)=\text{max}_x |\sum_{i=1}^M w_i u(x,\xi_i)-I^{\text{an}}(x)|$
    for the solution~\eqref{eq:transpnscatt}, with
    $\mathcal{R}(\xi)=0$. The errors are compared for the quadrature
    of Eq.~\eqref{eq:MartensenKussmaul}, and the Gauss--Legendre
    (G--L) quadrature.}
 \label{fig:intconvs}
\end{figure}
A suitable value $p=3.2$ was used, which provides excellent
convergence (Fig.~\ref{fig:intconvs}) for the integral in the $\xi$
variable while limiting the sharpness of the numerical boundary layer
in the $x$ variable. The latter boundary layer is subsequently
resolved by incorporating the logarithmic change of variables
$v=\log(\frac{x}{1-x})$, thus giving rise to points $x$ extremely
close to the boundary (without detriment to the integration process,
in view of the $r$-derivative bounds mentioned above, which are
uniform for all $x$), thus leading to high order precision in both the
$\xi$ and $x$ variables.

The transport equation is solved in a computational spatial domain
$[v_{\text{min}},v_{\text{max}}]$, with
$[x'_{\text{min}},x'_{\text{max}}]=[\frac{e^{v_{\text{min}}}}{e^{v_{\text{min}}}+1},\frac{e^{v_{\text{max}}}}{e^{v_{\text{max}}}+1}]$,
and with boundary conditions at $x'_{\text{min}}$ and
$x'_{\text{max}}$ obtained by means of the asymptotic boundary layer
solution~\eqref{eq:intf}.  By using the new variables the time
dependent transport problem
\begin{equation}
\begin{split}
\frac{\partial}{\partial t}\uvi+&\xi(2+2\cosh(v)) \frac{\partial}{\partial v}\uvi+\\
&\st\uvi=\sct \int_{-1}^{1} \uvp d\xi'+q, \\
u(v,\xi,t_{\text{min}})&=0,\\
u(v_{\text{min}},\xi,t)&=u_{0}(x'_{\text{min}},\xi,t) \; \forall \xi>0,\\
u(v_{\text{max}},\xi,t)&=u_{0}(x'_{\text{max}},\xi,t) \; \forall \xi<0.
\end{split}
\label{eq:transptd}
\end{equation}
results. The time propagation is performed implicitly by means of a third order backward 
differentiation formula (BDF). 
The collisional term is obtained by a polynomial extrapolation 
to avoid the inversion of large matrices at each time step. 
The discrete version of Eq.~\eqref{eq:transptd} reads
\begin{equation}
\begin{aligned}
& \left[ \hat{\mathds{1}} + \beta\Delta t\xi_j (2+2\cosh(v)) \hat{\mathds{D}} + \beta\Delta t \mu_t \hat{\mathds{1}} \right]  u^{n+1}_j  = \\
&\sum_{k=0}^{s-1}\alpha_k u_j^{n-k} +  \beta\Delta t \sct
 \sum_{i=1}^{M} w_i \tilde{u}_i^{n+1} + \beta\Delta t q^{n+1},
\end{aligned}
\label{eq:transptd1disc}
\end{equation}
with $u^{n+1}_j=u(v,\xi_j,t^{n+1})$, $t^{n+1}=n\Delta t$, where $\alpha_k$ and $\beta$ are the coefficients for the third 
order BDF formula, with $s=3$. The third order extrapolated quantity is given by 
$\tilde{u}_j^{n+1}=\sum_{\kappa=0}^{2}(-1)^\kappa {3 \choose \kappa+1} u_j^{n-\kappa}$. 
The operator $\hat{\mathds{1}}$ is the identity operator while $\hat{\mathds{D}}$ 
is a spectral differentiation operator, which yields an implicit version of the 
Fourier Continuation--Discrete Ordinates Method (FC--DOM) developed by the authors \cite{Gaggioli2019a}.

\begin{figure}[h!]
\centering
  \includegraphics[width=0.85\linewidth]{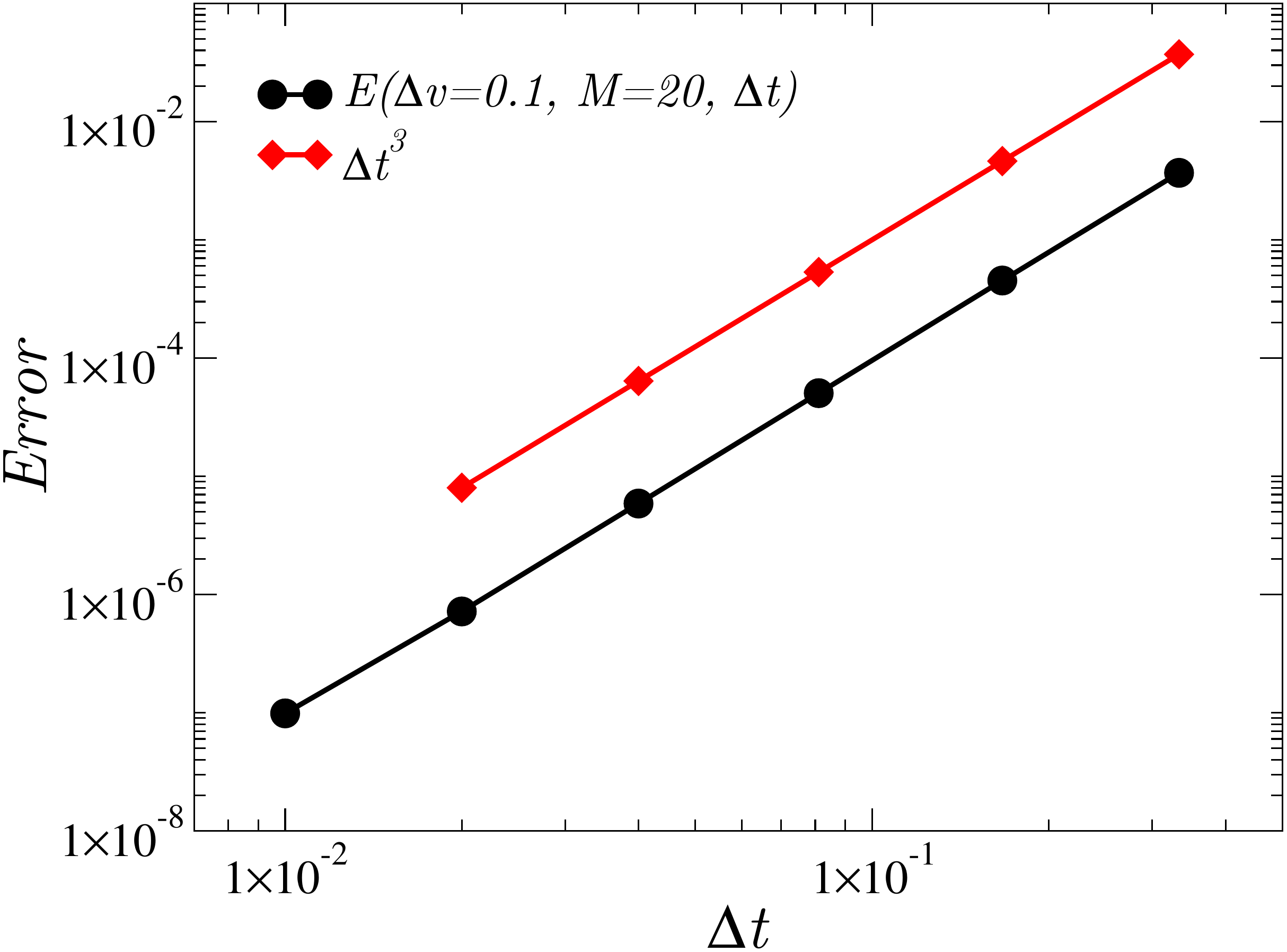}\\
    \vspace{0.3cm}
  \includegraphics[width=0.85\linewidth]{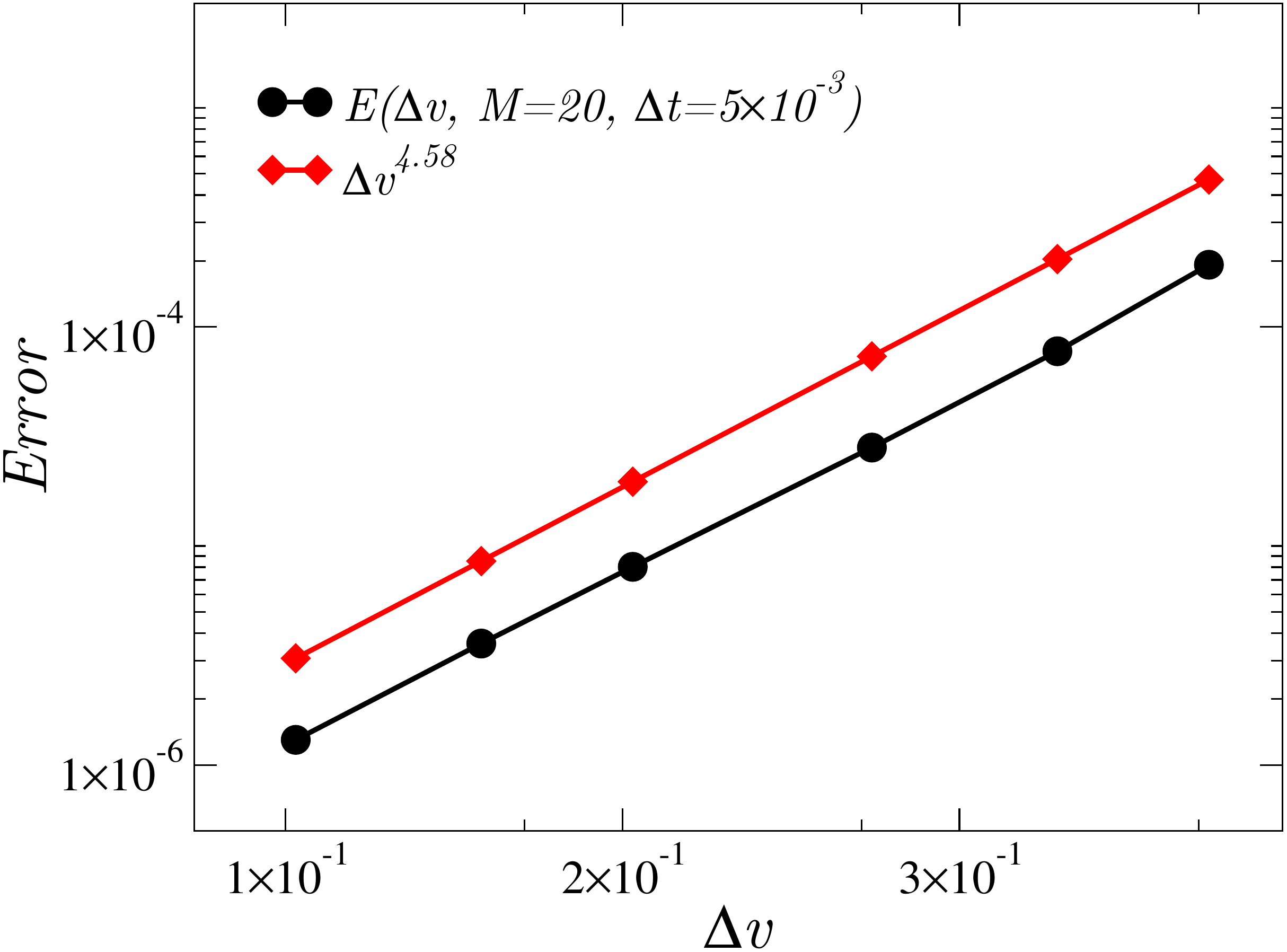}\\
    \vspace{0.3cm}
  \includegraphics[width=0.85\linewidth]{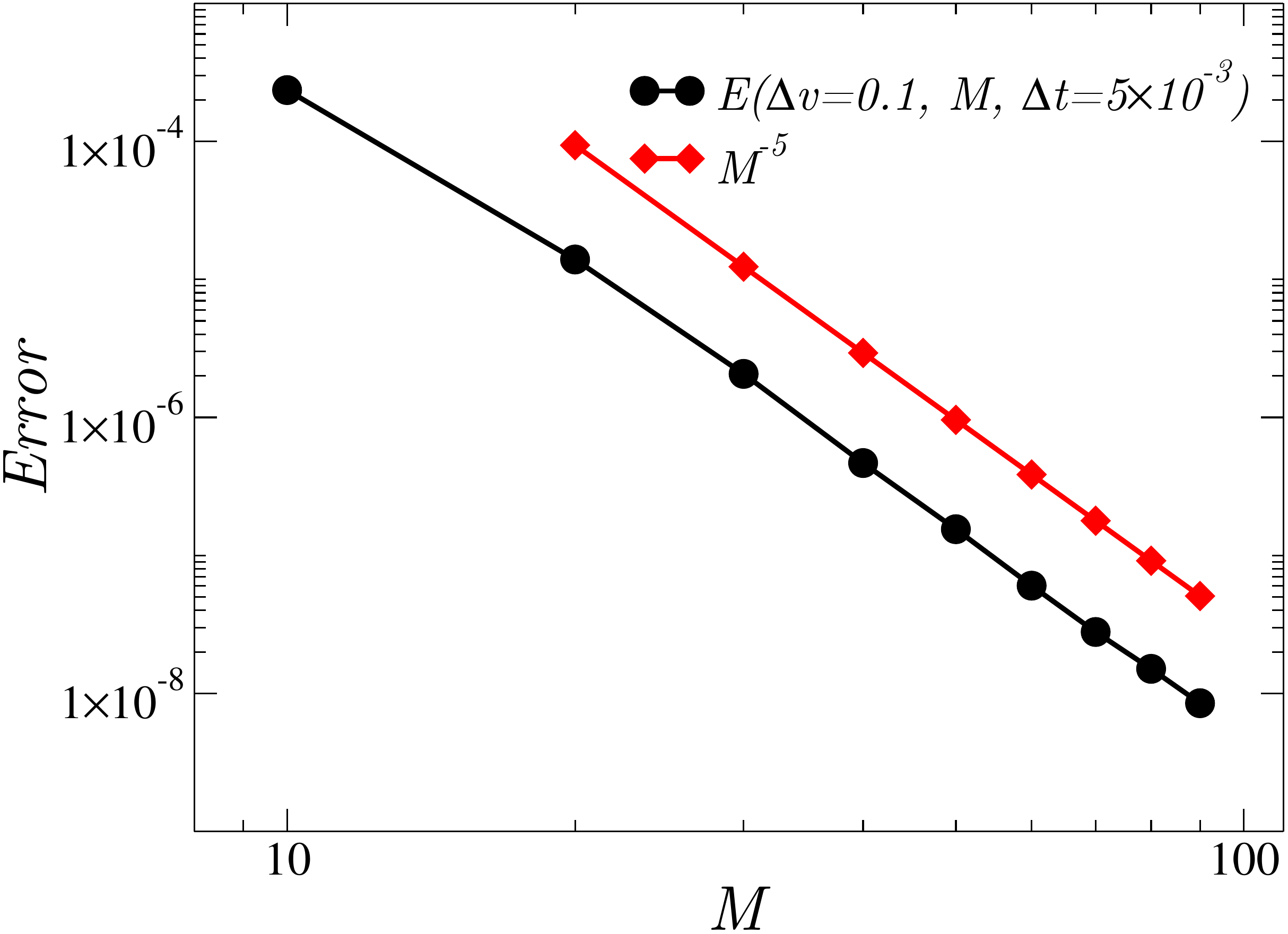}
  \caption{Convergence properties of the proposed algorithm, for all
    variables involved. In circles, the error
    $E(\Delta v, M,\Delta t)=\text{max}_{x,\xi}
    |u(x,\xi)-u^{\text{c}}(x,\xi)|$ is diplayed for various grids,
    where $u^{\text{c}}(x,\xi)$ denotes the converged solution.}
 \label{fig:conv}
\end{figure}
Fig.~\ref{fig:conv} demonstrates the excellent convergence properties of the algorithm, for $\mu_s=\mu_a=q=1$. Given that an analytic solution 
for problems including scattering is not known, the error was computed via comparison with the solution 
obtained on a finer grid. This high order of convergence clearly suggests that the changes of variables used in the $x$ and $\xi$ variables lead to adequate grid resolution of the boundary layers involved.

\begin{figure}[h!]
\centering
  \includegraphics[width=0.85\linewidth]{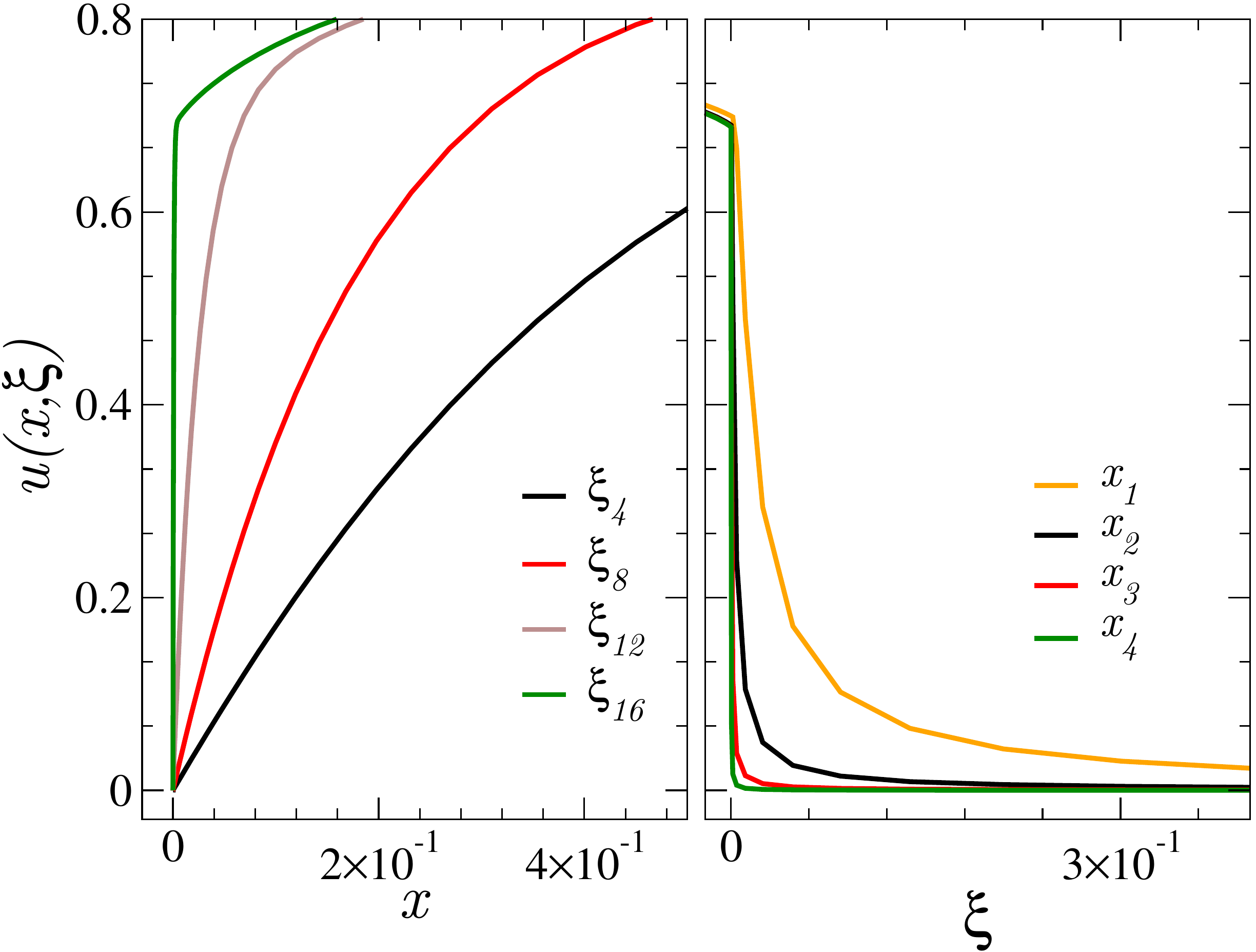}
  \caption{Boundary layers near $x=0$ obtained by solving Eq.~\eqref{eq:transptd}, with $\mu_a=\mu_s=q=1$ for different values of $\xi$ and $x$, with $\xi_i>\xi_{i+1}>0$
 and $x_i>x_{i+1}$.}
 \label{fig:blayers1}
\end{figure}

\begin{figure}[h!]
\centering
  \includegraphics[width=0.85\linewidth]{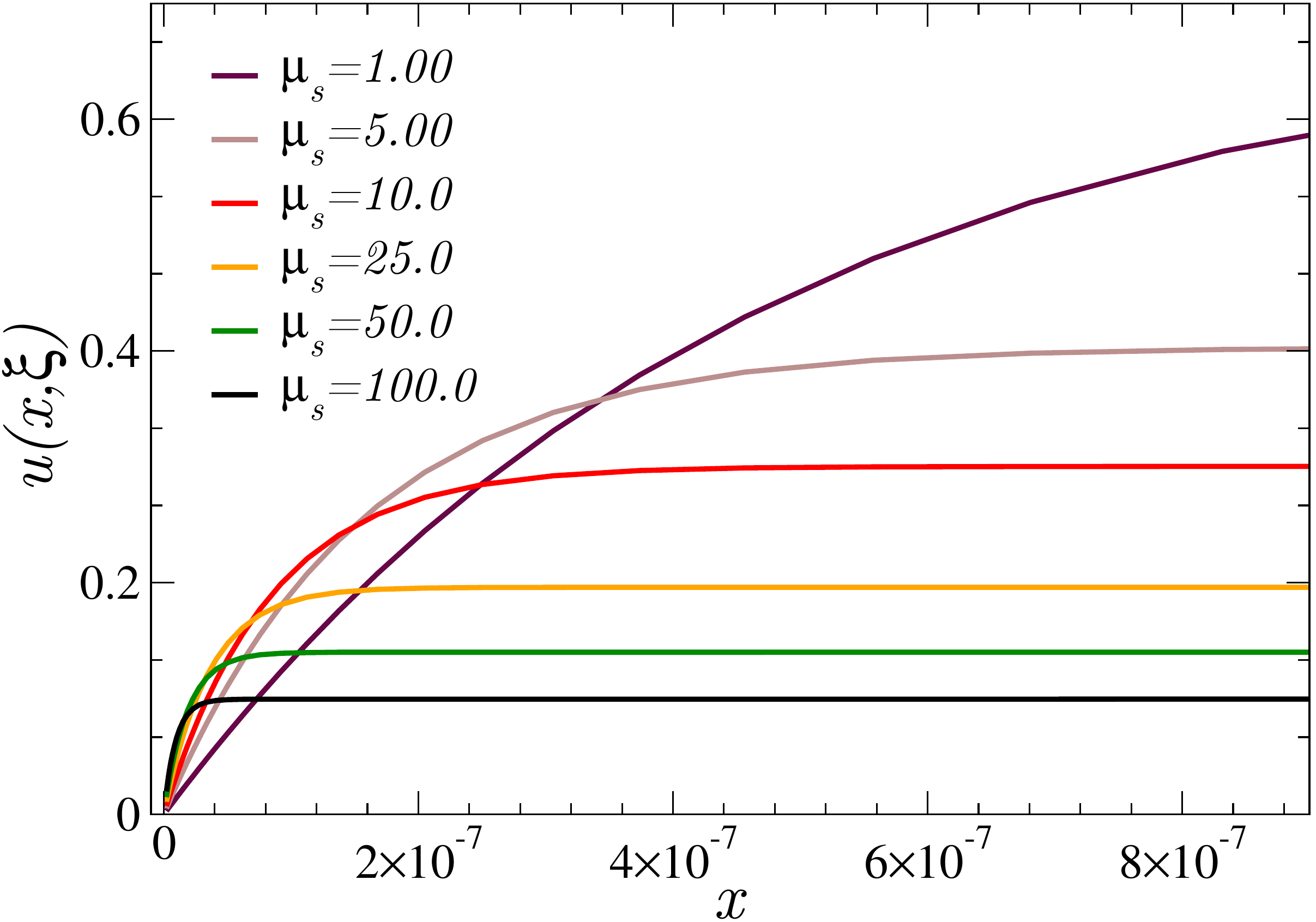}
  \caption{Boundary layers obtained from solving Eq.~\eqref{eq:transptd}, with $\mu_a=q=1$ 
  and different values of $\mu_s$. Solutions for direction with 
  $\xi=\xi_{\text{min}} \simeq 10^{-6}$ are shown. Note the large slopes of the 
  transport solution as $x\to 0^+$.}
 \label{fig:blayers}
\end{figure}

\begin{figure}[h!]
\centering
  \includegraphics[width=0.85\linewidth]{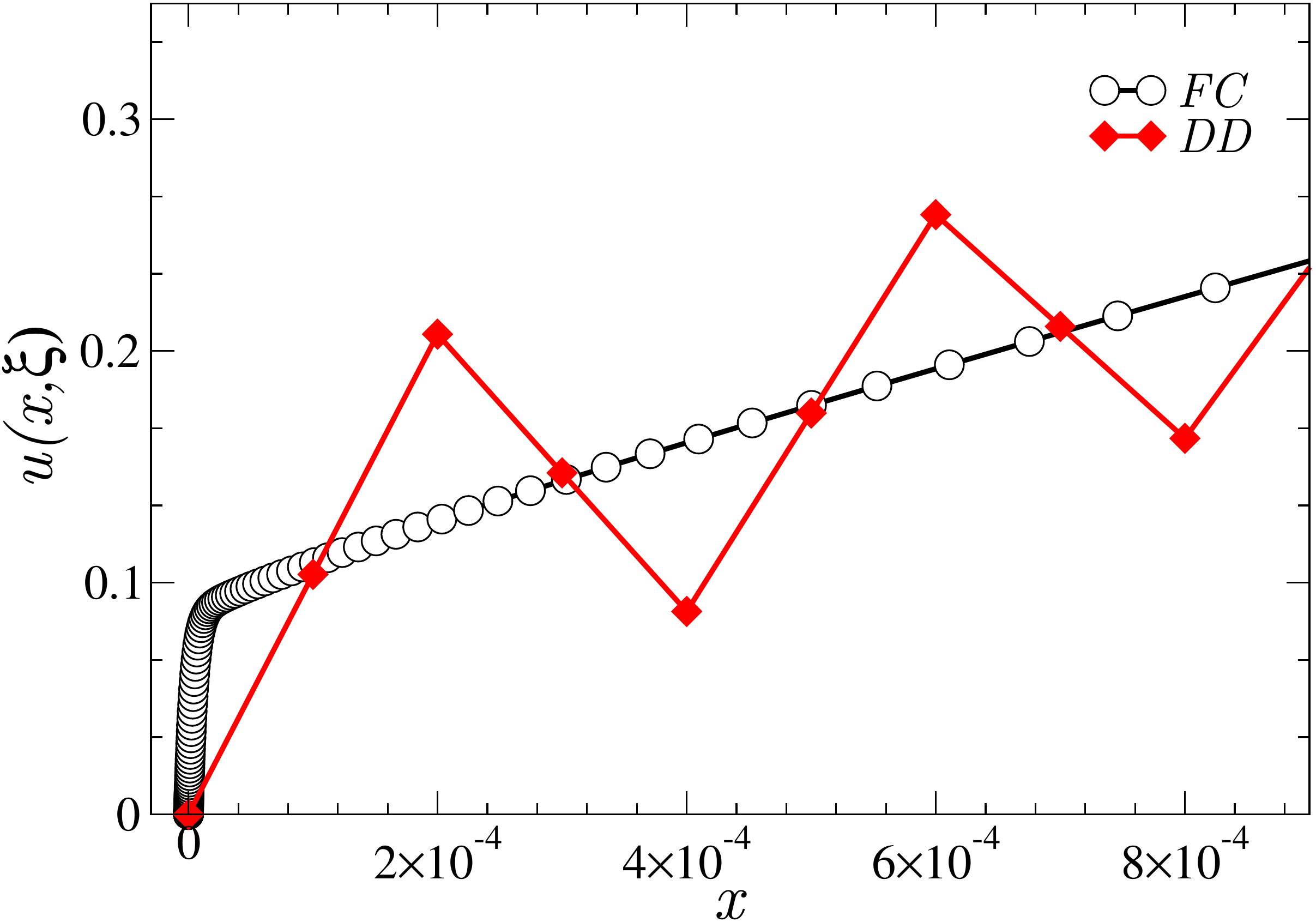}
  \caption{Solution $u(x,\xi_{15})$ of Eq.~\eqref{eq:transp1d} with 
$\mu_t=\mu_s=1000$, $q=0.1$ and $\xi_{\text{15}} \simeq 10^{-3}$, showing the oscillations originated in the DD scheme due to the large slopes at the boundary layer. 
$N=400$ discrete points are used in the spatial variable for the FC method, and $N=10000$ for the DD method.}
 \label{fig:DDlayer}
\end{figure}

In what follows the numerical algorithm is utilized to explore and
demonstrate the character of the boundary--layer structures considered
in this paper. For definiteness, in the rest of this paper we restrict attention
to time--independent solutions obtained by means of the time--dependent
solver and letting the solution relax in time, as described in
\cite{Gaggioli2019a}. The resulting steady--state solutions are
depicted in various forms in Figs.~\ref{fig:blayers1} through
\ref{fig:DDlayer}; similar boundary layer structures are of course
present for all times in the time-dependent solutions.

Fig.~\ref{fig:blayers1} presents boundary--layer structures in the $x$ and $\xi$ variables with $\mu_s=\mu_a=q=1$ (numerical parameter values $N=250$ and $M=40$ were used in these figures, 
with $-v_{\text{min}}=v_{\text{max}}=25$). As can be seen in Fig.~\ref{fig:blayers1}, the smaller 
values of $\xi$ give boundary layers that shrinks over smaller spatial 
regions, as expected from our boundary layer analysis. 

Fig.~\ref{fig:blayers} demonstrates the persistence of the boundary
layer even in presence of high scattering coefficient, with
$\mu_a=q=1$ fixed. The case $\xi=\xi_{\text{min}}\simeq 10^{-6}$ is
considered in the figure, with parameter values $N=200$, $M=20$ and
$-v_{\text{min}}=v_{\text{max}}=20$.  Clearly, even though diffusive
problems (large $\mu_s$) tend to be more regular over the $\xi$
variable---owing to the strong averaging and smoothing induced by the
large scattering coefficient---, the boundary layers that arise in the
spatial variable with increasing $\mu_s$ tend to have larger slopes as
$x\to 0^+$.

There have been many attempts to understand unphysical oscillations
associated with the widely used Diamond finite Difference scheme (DD)
for the transport equation~\cite{Larsen1987,Petrovic1996,Bal2001}.
The recent paper~\cite{Wang2019} avoids this problem by using only
directions $\xi$ away from
$\xi=0$. References~\cite{Larsen1987,Petrovic1996} attribute this type
of oscillations to anisotropic boundary conditions, non diffusive
boundary layers and/or high absorption; the present paper, which
demonstrates existence of boundary layers even in the isotropic case
and for all values of the scattering and absorption coefficients,
presents a starkly contrasting interpretation. In particular, the
present contribution explicitly demonstrates the existence of
exponential boundary layers, triggered by the boundary condition and
vanishing $\xi$ values, which are not considered in the previous
discussions.

For example, reference~\cite{Larsen1987} treats a diffusive transport
problem (Problem~1 in that reference) which, under rescaling, can be
reformulated as in Eq.~\eqref{eq:transp1d} with $\mu_s=\mu_t=1000$,
$q=0.1$ and $\mathcal{R}(\xi)=0$. This is an extremely diffusive
problem with isotropic boundary conditions for
which~\cite[pp. 317]{Larsen1987} states: ``...since the leading order
term in the asymptotic expansion of the analytic transport equation is
itself isotropic, this term in these problems does not contain a
boundary layer.'' In contrast, Fig.~\ref{fig:DDlayer} shows that
boundary layers are present in this problem.  The FC--DOM solution
displayed in this figure was obtained by means of $M=40$ discrete
directions and $N=400$ points in the spatial variable. In contrast,
$N=10000$ were used in our DD-scheme solution presented in
Fig.~\ref{fig:DDlayer}---which clearly displays the spurious
oscillations produced by the DD scheme in this context.

Physically, small values of $\xi$ along an incoming direction for
points $x$ near the boundary (cf. Fig.~\ref{fig:parallelgeom}) are
associated to long geometrical particle paths which are not present
for $x=0$, and which thus give rise to the sharp boundary-layer
transition described in this paper. We emphasize that this
boundary-layer structure, which was not previously reported in the
literature, constitutes a physical effect which was mispredicted for
nearly fifty years, and which, as demonstrated in
Fig.~\ref{fig:DDlayer} and throughout this paper, has a significant
impact on the physics and the numerical simulation of transport
phenomena.

OPB was supported by NSF through contract DMS-1714169 and by the
NSSEFF Vannevar Bush Fellowship under contract number
N00014-16-1-2808. ELG and DMM gratefully acknowledge the financial support from the following Argentine
institutions: Consejo Nacional de Investigaciones Cient\'{\i}ficas y
T\'ecnicas (CONICET), PIP 11220130100607, Agencia Nacional de Promoci\'on
Cient\'{\i}fica y Tecnol\'ogica (ANPCyT) PICT--2017--2945,
and Universidad de Buenos Aires UBACyT 20020170100727BA.


\end{document}